\documentclass[12pt]{article}
\usepackage{graphicx}
\usepackage{natbib}
\usepackage{url} 
\usepackage{amsthm,amsmath,amsfonts,amssymb}
\usepackage{mathtools}
\usepackage{xcolor}
\usepackage{enumerate}
\usepackage{multirow,booktabs,adjustbox,makecell,xspace}
\usepackage{times}
\usepackage{bm}
\usepackage{algorithm}
\usepackage{algorithmic}
\newcommand{\blind}{0}

\newtheorem{theorem}{Theorem}
\newtheorem{assumption}[theorem]{Assumption}
\theoremstyle{remark}
\newtheorem{remark}{Remark}%
\theoremstyle{definition}
\newtheorem{definition}{Definition}

\newcommand{\argmin}{\operatornamewithlimits{arg\ min}}
\newcommand{\argmax}{\operatornamewithlimits{arg\ max}}
\newcommand{\toP}{\overset{P}{\longrightarrow}}
\newcommand{\toD}{\overset{\mathcal D}{\longrightarrow}}

\newcommand{\bA}{{\mbox{\boldmath $A$}}}
\newcommand{\bB}{{\mbox{\boldmath $B$}}}

\newcommand{\bH}{{\mbox{\boldmath $H$}}}
\newcommand{\bI}{{\mbox{\boldmath $I$}}}
\newcommand{\bL}{{\mbox{\boldmath $L$}}}
\newcommand{\bM}{{\mbox{\boldmath $M$}}}
\newcommand{\bT}{{\mbox{\boldmath $T$}}}
\newcommand{\bR}{{\mbox{\boldmath $R$}}}

\newcommand{\bV}{{\mbox{\boldmath $V$}}}

\newcommand{\bX}{{\mbox{\boldmath $X$}}}
\newcommand{\bx}{{\mbox{\boldmath $x$}}}
\newcommand{\bY}{{\mbox{\boldmath $Y$}}}
\newcommand{\by}{{\mbox{\boldmath $y$}}}

\newcommand{\bu}{{\mbox{\boldmath $u$}}}
\newcommand{\bU}{{\mbox{\boldmath $U$}}}

\newcommand{\bS}{{\mbox{\boldmath $S$}}}
\newcommand{\bd}{{\mbox{\boldmath $d$}}}
\newcommand{\bG}{{\mbox{\boldmath $G$}}}

\newcommand{\iid}{i.i.d.\xspace}

\newcommand{\balpha}{{\mbox{\boldmath $\alpha$}}}
\newcommand{\bbeta}{{\mbox{\boldmath $\beta$}}}
\newcommand{\bmeta}{{\mbox{\boldmath $\eta$}}}

\newcommand{\btheta}{{\mbox{\boldmath $\theta$}}}

\newcommand{\AutoGFI}{{AutoGFI}\xspace}
\newcommand{\hessian}{{\bm{H}(\btheta^*)}}

\begin{document}

\def\spacingset#1{\renewcommand{\baselinestretch}%
{#1}\small\normalsize} \spacingset{1}



\if0\blind
{
\title{\bf AutoGFI: Streamlined Generalized Fiducial Inference for Modern Inference Problems in Models with Additive Errors}
  \author{Wei Du\thanks{
  The authors are in alphabetical order. They gratefully acknowledge partial support by the National Science Foundation under grants \textit{CCF-1934568}, \textit{DMS-1916125}, \textit{DMS-2113605}, \textit{DMS-2113404}, \textit{DMS-2210337}, and \textit{DMS-2210388}}\hspace{.2cm}\\
    Department of Statistics, University of California, Davis\\
    Jan Hannig \\
    Department of Statistics and Operations Research, \\
    University of North Carolina at Chapel Hill\\
    Thomas C. M. Lee \\
    Department of Statistics, University of California, Davis\\
    Yi Su \\
    Department of Statistics, University of California, Davis\\
    Chunzhe Zhang \\
    Department of Statistics, University of California, Davis}
  \maketitle
} \fi

\if1\blind
{
  \bigskip
  \bigskip
  \bigskip
  \begin{center}
    {\LARGE\bf Title}
\end{center}
  \medskip
} \fi

\vspace*{-0.9cm}
\begin{abstract}

The concept of fiducial inference was introduced by R. A. Fisher in the 1930s to address the perceived limitations of Bayesian inference, particularly the need for subjective prior distributions in cases with limited prior information. However, Fisher’s fiducial approach lost favor due to complications, especially in multi-parameter problems. With renewed interest in fiducial inference in the 2000s, generalized fiducial inference (GFI) emerged as a promising extension of Fisher’s ideas, offering new solutions for complex inference challenges. Despite its potential, GFI's adoption has been hindered by demanding mathematical derivations and complex implementation requirements, such as Markov Chain Monte Carlo (MCMC) algorithms. This paper introduces \AutoGFI, a streamlined variant of GFI designed to simplify its application across various inference problems with additive noise. \AutoGFI’s accessibility lies in its simplicity—requiring only a fitting routine—making it a feasible option for a wider range of researchers and practitioners. To demonstrate its efficacy, \AutoGFI is applied to three challenging problems: tensor regression, matrix completion, and network cohesion regression. These case studies showcase \AutoGFI’s competitive performance against specialized solutions, highlighting its potential to broaden the application of GFI in practical domains, ultimately enriching the statistical inference toolkit.
\end{abstract}

\noindent%
\vfill

\spacingset{1.75} 

\section{Introduction}\label{sec:Intro}
Fiducial inference was originally proposed by Fisher \cite{fisher1922mathematical,fisher1925theory,fisher1930inverse,fisher1933concepts,fisher1935fiducial} with the goal of constructing a distribution for parameters of interest. This so-called fiducial distribution can then be used for statistical inferences, such as constructing confidence sets. Like the Bayesian posterior distribution, the fiducial distribution is data-dependent, but the key distinction is that the fiducial approach does not demand a priori information about the parameter. Fisher showed that in simple settings, especially for one-parameter families of distributions, fiducial intervals coincide with classical confidence intervals. In multiple-parameter families of distributions, the fiducial distribution provides confidence sets whose coverage is close to the target confidence levels. However, controversies arose because, in multi-parameter settings, fiducial inference often led to procedures that were not exact in the frequentist sense. Also, there is often no unique way to define a fiducial distribution. Interested readers can find a detailed discussion on the controversies regarding fiducial inference in \cite{zabell1992ra}. Because of the non-exactness and non-uniqueness of the fiducial distributions, fiducial inference was not widely accepted among mainstream statisticians until its recent reincarnation in the 2000s.

In the past two decades, there has been a resurgence of interest in modern modifications of fiducial inference. These works include Dempster-Shafer theory \citep{Edlefsen_2009} and inferential models \citep{martin2010dempster,zhang2011dempster,martin2013inferential}, which focus on posterior probabilistic inferences without using priors. Another category of methods called confidence distributions \citep{singh2007confidence,xie2011confidence,xie2013confidence} seeks inferentially meaningful distributions on the parameter from a frequentist point of view. Objective Bayesian inference, on the other hand, uses model-based non-subjective priors under the Bayesian framework. More recently, \cite{XiePeng2022} combine ideas from confidence distributions and inferential models to obtain an algorithmic-based approach to inference. 
The latest addition is the highly scalable and effective extended fiducial inference (EFI) of \cite{EFI2024}, which uses stochastic gradient Markov chain Monte Carlo to impute random errors in observations and employs a sparse deep neural network to estimate inverse functions.
A common thread of these approaches is to obtain some inferentially meaningful probability statements about the parameter space without the need for subjective prior information.

Despite all these efforts, generalized fiducial inference (GFI) \citep{Jan16}, another modification of fiducial inference, still has its edge in many areas. For example, GFI often offers good alternatives in terms of both performance and usability, e.g., the generalized fiducial distribution (GFD), which plays a similar role as the posterior distribution in the Bayesian context, is never improper. We believe that GFI and its quickly evolving variants have the potential to uncover profound and essential understandings of statistical inference. 

GFI has been applied to a variety of applications and shown promising results, such as 
wavelet regression \citep{Jan09b},
ultrahigh-dimensional regression \citep{Lai15}, 
binary response models \citep{Liu16}, 
exoplanet detection \citep{GFIspecline},
and many others \citep[e.g.,][]{
	mcnally2003tests,
	lidong2008fiducial}.
The theoretical properties of GFI have also been extensively studied using asymptotics in \cite{hannig2009generalized,Jan13,Jan06,majumder2016higher,Sonderegger14}.

Until now, the practical implementation of GFI has often required a complete or partial calculation of the GFD, which may involve pairing it with Markov Chain Monte Carlo (MCMC) techniques to generate samples, known as fiducial samples, from the parameter space. This process can be quite tedious or, in some cases, even impossible, which limits the attractiveness of GFI for practitioners. 
Furthermore, in overparametrized settings that are common for many modern applications, the use of GFI requires additional efforts to avoid suboptimal performance. It typically requires non-trivial, case-by-case implementation of regularization and debiasing procedures to ensure the effectiveness of the GFI solution.

{
To overcome these issues, this paper proposes an innovative approach designed to simplify and enhance the application of GFI under the additive noise setting.  It is called {\em \AutoGFI}, which is an algorithm that possesses the following properties:
\vspace{-0.2cm}
\begin{itemize}
\item {\em Accessibility}: \AutoGFI\ is engineered to be user-friendly and readily implementable.  Essentially, it can be implemented as long as a fitting routine is available.
\item {\em Simplification}: \AutoGFI\ eliminates the typically complex mathematical derivations and the need for intricate Markov Chain Monte Carlo (MCMC) algorithms associated with GFI. It can generate fiducial samples without the need for any analytical calculation of the generalized fiducial distribution. 
\item {\em Versatility}: \AutoGFI\ is versatile in its application and can be applied to a broad range of applications where no inferential tools were previously available.
\item {\em Performance}: \AutoGFI\ demonstrates highly competitive performance when applied to a variety of complex problems, showcasing its effectiveness in practice.
\item {\em Innovation}: \AutoGFI\ represents a novel approach to fiducial inference, offering a fresh perspective on how to address statistical inference challenges.
\end{itemize}
\vspace{-0.1cm}

Overall, the introduction of \AutoGFI provides a solution to the challenge of implementing GFI in complex practical applications, making it a more accessible, appealing, and viable option for researchers and practitioners to solve inference problems. 
}

The rest of this article is structured as follows: Section~\ref{sec:GFIBackground} provides background information on GFI. Sections~\ref{sec:NewGFI} and~\ref{sec:debias} introduce the basic form of \AutoGFI, as well as its regularized and debiased counterparts. The theoretical properties of \AutoGFI\ are then explored in Section~\ref{sec:theory}, followed by applications to tensor regression in Section~\ref{sec:TR}, matrix completion in Section~\ref{sec:MC}, and regression with network cohesion in Section~\ref{sec:NR} to demonstrate its wide applicability and excellent empirical performance. Concluding remarks are in Section~\ref{sec:conclude}, while technical details are deferred to the appendix.

\section{Background on Generalized Fiducial Inference}\label{sec:GFIBackground}
\subsection{A General form of GFI}\label{sec:FItoGFI}

The development of GFI is essentially inspired by our understanding of Fisher's fiducial argument. We will present it by linking it to the widely accepted likelihood function. Recall that $f(x;\theta)$ is the probability density function of a random variable $X$ when we treat $\theta$ as a fixed unknown parameter and $X$ as a random value. The likelihood function is obtained when we switch the role of the random variable and the parameter. In other words, given observed data $x$, the pdf $f(x;\theta)$ becomes the likelihood function, a function $l_x(\theta)$ of $\theta$. Our understanding of fiducial is also backed up by this ``role-switching'' mechanism; \cite{hannig2009generalized} contains a detailed example. 
Below we formally introduce this idea.

GFI starts with a data generating equation describing the relationship between the data $\bY$ and the parameter $\btheta$. The data generating equation can be written as
\begin{equation}\label{eqn:DGE}
\bY =\bm{F}(\btheta, \bU),
\end{equation}
where $\bm{F}$ is a deterministic function, and $\bU$ is the random component whose distribution is completely known. 

Now applying the ``role-switching'' idea as in likelihood function definition: for any observed data $\by$, 
we define the following ``inverse'' mapping of the data generating equation:
\begin{equation}\label{eqn:Q}
\bm{Q}_{\by}(\bu) = \argmin_{\btheta} \rho\left(\bm{F}(\btheta, \bu), \by\right),
\end{equation}
where $\rho$ is a smooth semi-metric, e.g., squared $\ell_2$ norm $\rho(\by,\by^*)=\\frac{1}{2}\|\by-\by^*\|^2_2$. 
Given a realization $\bu$ of $\bU$, $\bm{Q}_{\by}(\bu)$ is a value of parameter $\btheta$ such that $\bm{F}(\btheta, \bu)$ comes closest to the observed data~$\by$.

When the inverse (in $\btheta$) of \eqref{eqn:DGE} exists then
\begin{equation}\label{eqr:inverse}
\by=\bm{F}(\bm{Q}_{\by}(\bu), \bu).
\end{equation}
However, for many complex statistical problems the exact inverse property \eqref{eqr:inverse} cannot be guaranteed for all $\by$ and $\bu$. However, even then, there must exist some $\bu$ for which the equality \eqref{eqr:inverse} holds, as we assume that the data $\by$ was generated using \eqref{eqn:DGE}. Denote the set of all such $\bu$ as $\mathcal U_{\by,0}$.

Heuristically speaking,  a sample of $\{\btheta_i^*\}_{i=1}^m$ from the GFD is obtained by first generating a series of independent $\{\bU_i^*\}_{i=1}^{m}$ from $\bU$'s distribution truncated to $\mathcal U_{\by,0}$, i.e., generating $\bU_i^*$ conditional on the event that they fall into $\mathcal U_{\by,0}$, and then setting $\btheta_i^*=\bm{Q}_{\by}(\bU_i*)$. Due to Borel paradox \citep[sec. 4.9.3]{casella2002statistical}, the above truncated distribution is ill-defined when $P(\bU\in\mathcal U_{\by,0})=0$. To address this, we enlarge this set by adding a small tolerance, defining $\mathcal{U}_{\by,\epsilon}:=\{ \bu: \rho( \by,\bm{F}(\btheta,\bu)) \leq \epsilon\}$. As this tolerance vanishes, it leads to the following limit definition of GFD \citep{Jan16}.
\begin{definition}\label{def:limitGFI}
    Let $\bU^*_\epsilon$ follows the distribution of $\bU$ truncated to $\mathcal U_{\by,\epsilon}$, i.e., having density 
    
    \noindent $f_{\bU}(\bu)I_{\mathcal{U}_{\by,\epsilon}}(\bu)/P(\bU\in\mathcal U_{\by,\epsilon})$, where $f_{\bU}(\bu)$ is the density of $\bU$.  Denote the distribution of $\bm{Q}_{\by}(\bU^*_\epsilon)$ by $\mu_\epsilon$. If the \textit{weak limit}
    $\lim_{\epsilon\rightarrow 0} \mu_\epsilon$ exists,
    the limit is called a generalized fiducial distribution (GFD).
\end{definition}

In practice one could select a small $\epsilon>0$ and generate an approximate fiducial sample $\btheta^*=\bm{Q}_{\by}(\bU^*_\epsilon)$.
We note that this is very similar to Approximate Bayesian Computations \citep[ABC,][]{beaumont2002approximate}. While both methods compare $\by^*=\bm{F}(\btheta^*, \bU^*)$ with the observed data $\by$, the main difference is that ABC generates $\btheta^*$ from a prior distribution while GFD uses the best-fitting $\btheta^*$ obtained from the optimization problem \eqref{eqn:Q}. This observation provides an interesting philosophical connection between fiducial and Bayesian inference that we plan to investigate in future work.

Finally, we remark that if multiple solutions $\bm{Q}_{\by}(\bu)$ exist, one can simply select one of them using a possibly random rule. Some guidance of such selection can be found in \cite{Jan13}.  In fact, the uncertainty due to multiple solutions will only introduce a second-order effect on the statistical inference in many parametric problems \citep{Jan16}. 

\subsection{The Jacobian Formula}\label{sec:jacob}
Under some smoothness assumptions, 
\cite{Jan16} show
that the limiting distribution in Definition~\ref{def:limitGFI} has a density
\begin{equation}\label{eqn:GFD}
r(\btheta|\by) = \frac{f(\by,\btheta)J(\by,\btheta)}{\int_{\Theta} f(\by,\btheta')J(\by,\btheta') d\btheta'},
\end{equation}
where $f(\by,\btheta)$ is the likelihood function, and 
\begin{equation}\label{eqn:GFDjacob}
J(\by,\btheta) = D\left( \nabla_{\btheta}\bm{F}(\bu,\btheta) \bigg|_{\bu=\bm{F}^{-1}(\by,\btheta)} \right).
\end{equation}
When using squared $\ell_2$ norm as $\rho$ in (\ref{eqn:Q}), it is shown in \cite{Jan16} that $D(\nabla_{\btheta}\bm{F})=(\det \nabla_{\btheta}\bm{F}^\top\nabla_{\btheta}\bm{F})^{1/2}$. Also, $\bu=\bm{F}^{-1}(\by,\btheta)$ is the value of $\bu$ such that $\by = \bm{F}(\btheta, \bu)$.

Equations~(\ref{eqn:GFD}) and~(\ref{eqn:GFDjacob}) present an interesting and intriguing connection between GFI and Bayesian methodology: the density $r(\btheta|\by)$ in~(\ref{eqn:GFD}) behaves like a posterior density in the Bayesian context with $J(\by,\btheta)$ being the ``prior'', except that the data $\by$ also appear in $J(\by,\btheta)$, so strictly speaking it is not a prior density.  Note also that $J(\by,\btheta)$ shares the invariance to reparametrization property with the Jeffreys prior.

When using~(\ref{eqn:GFD}) and~(\ref{eqn:GFDjacob}) for GFI applications, typically there are three possibilities:
\begin{enumerate}
	\item a closed form expression for $r(\btheta|\by)$ can be obtained,
	\item $r(\btheta|\by)$ is known up to a normalizing constant, and
	\item the term $J(\by,\btheta)$ cannot be analytically calculated so (\ref{eqn:GFD}) cannot be applied.
\end{enumerate}
The first possibility often happens only for simple problems where alternative inference solutions are already known.  For the second possibility, MCMC methods are required to generate a fiducial sample from $r(\btheta|\by)$, which could be computationally demanding.  
Therefore, the Jacobian formula is not always practical. More complex problems belong to the third possibility because $J(\by,\btheta)$ is not straightforward to work with. To overcome this issue, we work from the definition of GFD (Definition~\ref{def:limitGFI}) and propose a new form that is friendly to practitioners and easy to use in many modern applications.
\section{\AutoGFI for Additive Noise Models}\label{sec:NewGFI}
Now we focus on data generating equations of the form $\bY = \bm{G}(\bX,\btheta)+\bU$, where $\bU=(U_1,\ldots,U_n)^\top$ has a known distribution, e.g. $U_i$ are \iid $N(0,\sigma^2)$.  
Many popular statistical and machine learning models are of this kind.  For computational reasons, it is often advisable to treat the noise variance $\sigma^2$ as known and then use a consistent plug-in estimator of $\sigma^2$ in applications.  First, we look at the following algorithm that approximately generates samples from the GFD described at the end of Section~\ref{sec:FItoGFI}:
\begin{enumerate}
	\item Generate an independent copy $\bU^*$ of $\bU$.
	\item Solve $\btheta^* = \argmin\limits_{\btheta} \rho(\bY,\bm{G}(\bX,\btheta)+ \bU^*)$.
	\item Accept $\btheta^*$ if $\rho(\bY,\bm{G}(\bX,\btheta^*)+ \bU^*) \leq\epsilon$; otherwise reject and return to Step 1.
\end{enumerate}
This algorithm can be viewed as a fiducial version of approximate Bayesian computation \citep[ABC,][]{beaumont2002approximate}.
Typically one sets $\epsilon=\epsilon' \delta_0$, where $\delta_0= \min\limits_{\btheta} \rho(\bY,\bm{G}(\bX,\btheta))$ and $0<\epsilon'<1$. For certain problems the optimization in Step 2 can be solved in a closed form, making the algorithm very easy to implement.


We propose to generalize this basic GFI algorithm in the following ways. First, a penalty term $\lambda(\btheta)$ is added to Step~2 to introduce regularization or shrinkage to handle, say, model selection problems.  Next, when shrinkage is applied, we also propose including a debiasing operation $\bm{d}(\btheta)$ to the algorithm. We will provide a general approach to finding such a debiasing function in Section~\ref{sec:debias}.
Finally, theoretical results in Section~\ref{sec:theory} show that the resulting inference is valid even if the rejection Step~3 is omitted. Therefore, in order to speed up the calculation, we choose a relatively large $\epsilon$, excluding only a small sample that appears to be outliers.
The resulting easy-to-use  \AutoGFI algorithm is summarized in Algorithm~\ref{alg:AutoGFI}. We note that this algorithm can be applied to a wide range of modern applications as long as there is a fitting procedure for the parameters, i.e., Step~2 can be executed. 


\begin{algorithm}
\caption{\AutoGFI} \label{alg:AutoGFI}
\textbf{Input:} Data: $\bX$, $\bY$ \\
\textbf{Output:} Debiased fiducial samples $\btheta^*_\text{de}$ for the parameter $\btheta$ 
\begin{enumerate}
\item Generate an independent copy $\bU^*$ of $\bU$.
\item Solve $\btheta^* = \argmin\limits_{\btheta} \rho(\bY,\bm{G}(\bX,\btheta)+\bU^*) + \lambda(\btheta)$.
\item Debias $\btheta^*$ with $\btheta^*_\text{de} = \bd(\btheta^*)$.
\item Accept $\btheta^*_\text{de}$ if $\rho(\bY ,\bm{G}(\bX,\btheta^*_\text{de})+\bU^*) \leq\epsilon$; otherwise reject and return to Step 1.
\end{enumerate}
\end{algorithm}


Here we highlight the distinction between \AutoGFI\ and the celebrated bootstrap \citep{Efron1994bootstrap}. Non-parametric bootstrap is based on the ``re-sampling'' idea, such as re-sampling pairs in regression setup. In contrast, parametric bootstrap must rely on an ``initial'' fitted model, and hence, the inference made is sensitive to the initial fitting. On the other hand, the key of \AutoGFI\ is re-sampling the random component $\bU$ acting similarly to a ``pivot'', as its distribution does not depend on the parameters, and then refitting the perturbed data as in Step 2 of the algorithm. Recall that we recommend selecting a large value of $\epsilon$ in Step~4 of Algorithm~\ref{alg:AutoGFI}, and so the effect of the rejection step on the computational efficiency of \AutoGFI is minimal. Reader interested in exploring how GFI is related to pivots and related notions is referred to \cite{CuiHannig2024}.


\section{Debiasing for \AutoGFI}\label{sec:debias}
As hinted before, if a penalty term is added in Step~2, the \AutoGFI\ estimates necessarily suffer from non-negligible bias.  Inspired by the idea proposed in \cite{van2014asymptotically,jankova2018inference,WangEtAl2021}, 
this section provides a general approach to correct such estimation bias.


Under differentiability conditions, a potential fiducial sample generated by Step 2 can often be viewed as solving the estimating equation 
\begin{equation}\label{eq:subgradiant}
    \nabla^\top_\btheta \rho(\bY,\bm{G}(\bm{X},\bm{\theta})+ \bU^*)|_{\btheta=\btheta^*} + \bm{\xi}(\btheta^*)=0,
\end{equation}
where $\bm{\xi}(\btheta^*)$ is a (sub-)gradient of the penalty term $\lambda(\bm{\theta})$ evaluated at $\btheta^*$. The idea of removing the bias associated with the penalty is to modify the fiducial sample $\btheta^*$ so that the first term of \eqref{eq:subgradiant} is closer to 0. To this end, we use a one-step modification and define the debiased fiducial sample $\btheta^*_\text{de}$ by solving the equation
\begin{equation}\label{eq:approxDebias}
    -\bm{\xi}(\btheta^*)
    + \hessian
    (\btheta^*_\text{de}-\btheta^*) = \bm{0},
\end{equation}
where $\hessian$ is the Hessian matrix of second partial derivatives  of $\rho(\bY,\bm{G}(\bm{X},\bm{\theta})+\bU^*)$ with respect to $\btheta$ evaluated at ${\btheta=\btheta^*}$. An important exception is when coordinate $j$ of the subgradient ${\xi}(\btheta^*)_j$ is an interval containing 0. In that case, we do not debias the $j$th coordinate of the fiducial sample: set $\theta_{j,\text{de}}^*=\theta_j^*$ and remove the corresponding coordinates in \eqref{eq:approxDebias}, and also remove the appropriate rows and columns from the matrix $\hessian$. This exception assures that our debiasing procedure does not interfere with model selection. For example, if an $\ell_1$ penalty shrinks some coefficients to 0, the subgradient corresponding to these coefficients will be an interval containing 0, and by the proposed rule, these coefficients will be kept as~0 \citep{Boyd2004Convex}.

In high-dimensional settings, the matrix $\hessian$ is usually rank-deficient and poorly conditioned. As a result, the solution to \eqref{eq:approxDebias} is numerically unstable and highly variable. 
First, when penalty-inducing sparsity is used, we control the variability by treating Step 2 as a model selection step and then only performing debiasing for the non-zero coordinates in which $\xi(\btheta^*)$ is not an interval containing 0. 
Second, we suppress this variability by using a pseudo-inverse $\hessian^{\text{pinv}}$ instead of the real inverse $\hessian^{-1}$. 

For a square matrix $\bH \in \mathbb{R}^{n \times n}$ with SVD as $\bH = \bV \bm{\Sigma} \bm{W}^\top$, where $\bm{\Sigma}$ is the diagonal matrix containing all non-zero singular values, denote by $\zeta_i(\bH)$ the $i$-th largest singular value of $\bH$. Let $\bm{S}$ be the diagonal matrix containing all the singular values greater than $c\,\zeta_1(\bH)$ for some threshold constant $c$. We define the pseudo inverse of $\bH$ as
$$
\bH^{\text{pinv}} := \bm{W} \begin{pmatrix}
    \bm{S}^{-1} & 0 \\
    0 & 0
\end{pmatrix}
\bV^\top.
$$ 
In other words, we only use those singular vectors corresponding to significant singular values of $\hessian$ to perform debiasing. The threshold $c$ can be determined in a data-dependent manner. For example, $c$ can be chosen as any value between $\zeta_i(\bH)/\zeta_{1}(\bH)$ and $\zeta_{i+1}(\bH)/\zeta_{1}(\bH)$ where $i = \argmax\limits_k \zeta_{k}(\bH)/\zeta_{k+1}(\bH)$. This implies that $c$ is located at the point where there is a large jump in the magnitude of the singular values of $\bH$. With these, the debiasing function is defined as
\begin{equation}\label{eq:deBiasPin}
\btheta^*_{\text{de}} :=\btheta^*+\hessian^{\text{pinv}}\bm{\xi}(\btheta^*).
\end{equation}


The remaining problem is calculating the first and second derivatives of $\rho$, which could be challenging for complex models. To tackle this, we advocate using a set of new techniques in mathematics and computer algebra called {\em automatic differentiation}.  With such techniques, we can accurately and efficiently evaluate the derivative of a function specified by a computer program.  Many packages have been developed to implement these techniques.  In our work, we use the {\tt Python} package {\tt JAX} \citep{frostig2018compiling} to carry out the calculations.  

\section{Theoretical Results}\label{sec:theory}
This section presents some theoretical properties of \AutoGFI. All limits in this section are taken as $n\to\infty$. The proofs are delayed to the appendix.

Let $Y_i=G(\bX_i,\btheta_0)+ U_i$ for all $i=1,\ldots,n$ and $\bU^*$ is an independent copy of $\bU=(U_1,\ldots,U_n)^\top$, i.e., $\bU$ and $\bU^*$ are \iid Assume $\rho(\bx,\by)=\|\bx-\by\|^2/2=\sum_{i=1}^n(x_i-y_i)^2/2$, the gradient of the penalty function $\bm{\xi}_n(\btheta)$ is monotone increasing and differentiable function, and the data generating function $\bG(\bX,\btheta)$ is twice continuously differentiable in $\btheta$. Consequently \eqref{eq:subgradiant} becomes 
\[
\sum_{i=1}^n - \nabla^\top_\btheta G(\bX_i,\btheta)(Y_i-G(\bX_i,\btheta)- U_i^*)+\bm{\xi}_n(\btheta)=0.
\]

Denote by $\hat\btheta$ the solution of 
\begin{equation}\label{eq:PointEst}
\sum_{i=1}^n  - \nabla^\top_\btheta G(\bX_i,\hat\btheta)(Y_i-G(\bX_i,\hat\btheta))+\bm{\xi}_n(\hat\btheta)=0,
\end{equation}
by $\btheta^*$ the solution of 
\[
 \sum_{i=1}^n - \nabla^\top_\btheta G(\bX_i,\btheta^*)(Y_i-G(\bX_i,\btheta^*)- U_i^*)+\bm{\xi}_n(\btheta^*)=0,
\]
and by
$\hat\btheta_0$ the solution of 
\begin{equation}\label{eq:BiasEst}
 \sum_{i=1}^n  - \nabla^\top_\btheta G(\bX_i,\hat\btheta_0)(G(\bX_i,\btheta_0)-G(\bX_i,\hat\btheta_0))+\bm{\xi}_n(\hat\btheta_0)=0.
\end{equation}

Recall that all limits will be taken as $n\to\infty$. We assume the following:
\begin{assumption}\label{ass:consistency}
 There exists a compact $K\subset\Theta$ so that $\btheta_0\in K^o$ the interior of $K$.
 \begin{enumerate}[(a)]
     \item The probability that there exist $\hat\btheta_0\in K$, $\hat\btheta\in K$, and $\btheta^*\in K$ converges to 1.
     
     \item There exists a monotone increasing function $h$ so that $h(0)=0$ and for all $\btheta\in K$ 
      $$
       P \Bigl(\|n^{-1}\sum_{i=1}^n \nabla^\top_\btheta G(\bX_i,\btheta)(G(\bX_i,\btheta_0)-G(\bX_i,\btheta))\| 
        \geq h(\|\btheta-\btheta_0\|)\Bigr)\to 1.$$
      
   \item There exists a full rank, continuous {covariance matrix} $\bS(\btheta)$ so that 
   \[
     n^{-1/2}  \sum_{i=1}^n \nabla^\top_\btheta G(\bX_i,\btheta)  U_i \toD  N(\bm{0},\bS(\btheta))
    \]
    uniformly in $\btheta\in K$.
    \item The penalty satisfies $n^{-1}\sup_{\btheta\in K} \bm{\xi}_n(\btheta)\to 0$.
 \end{enumerate}
\end{assumption}

\begin{theorem}\label{thm:consistency}
 Under Assumption~\ref{ass:consistency},
$\|\hat\btheta-\btheta_0\|\toP 0$ and $\|\btheta^*-\btheta_0\|\toP 0$.
\end{theorem}

\begin{assumption}\label{ass:normality}
  Consider the Taylor series approximation at $\hat\btheta_0$
    \begin{multline}\label{eq:Taylor}
      n^{-1}\left(\sum_{i=1}^n  - \nabla^\top_\btheta G(\bX_i,\btheta)(G(\bX_i,\btheta_0)-G(\bX_i,\btheta))+\bm{\xi}_n(\btheta)\right)\\
      =\bT_n(\hat\btheta_0)(\btheta-\hat\btheta_0) + \bR_n(\btheta,\hat\btheta_0).
  \end{multline}
  Assume
  \begin{enumerate}[(a)]
      \item $\bT_n(\hat\btheta_0)\toP \bT_\infty$, where $ \bT_\infty$ is invertible.
      \item There is a continuous function $R$, so that $R(0)=0$, and $\tau>0$ 
      so that for all 
      $\|\btheta-\hat\btheta_0\|\leq\tau$,
      \[
       P\left(\|\bR_n(\btheta,\hat\btheta_0)\|\leq \|\btheta-\hat\btheta_0\|R(\|\btheta-\hat\btheta_0\|)\right)\to 1.
      \]
  \end{enumerate}
\end{assumption}
\begin{theorem}\label{thm:normality}
    Under the Assumptions~\ref{ass:consistency} and \ref{ass:normality},
    \[
    n^{1/2}(\hat\btheta-\hat\btheta_0)\toD N(\bm{0},\bT_\infty^{-1} \bS(\btheta_0)\bT_\infty^{-1\top} ),\]
     \[n^{1/2}(\btheta^*-\hat\btheta)\toD N(\bm{0},\bT_\infty^{-1} \bS(\btheta_0)\bT_\infty^{-1\top} ).
    \]
\end{theorem}

Notice that if $n^{-1/2}\bm{\xi}_n(\btheta)\to \bm{0}$, Theorem~\ref{ass:normality} implies that confidence intervals based on the generalized fiducial distribution will be asymptotically correct. Otherwise, recall the debiasing procedure \eqref{eq:deBiasPin} and notice that under assumptions of this section
\begin{equation*}
  \bH(\btheta^*)=
  \sum_{i=1}^n  - (\nabla_\btheta \nabla^\top_\btheta G(\bX_i,\btheta^*))(Y_i-G(\bX_i,\btheta^*)- U_i^*)
  +\nabla_\btheta G(\bX_i,\btheta^*)\nabla^\top_\btheta G(\bX_i,\btheta^*).
\end{equation*}

Next define
\begin{equation*}
 \hat\bH(\btheta)=
  \sum_{i=1}^n - (\nabla_\btheta \nabla^\top_\btheta G(\bX_i,\btheta))(Y_i-G(\bX_i,\btheta)) +
  \nabla_\btheta G(\bX_i,\btheta)\nabla^\top_\btheta G(\bX_i,\btheta),
\end{equation*}
\begin{equation*}
 \bH_0(\btheta)=
  \sum_{i=1}^n - (\nabla_\btheta \nabla^\top_\btheta G(\bX_i,\btheta))(G(\bX_i,\btheta_0)-G(\bX_i,\btheta))\\+
  \nabla_\btheta G(\bX_i,\btheta)\nabla^\top_\btheta G(\bX_i,\btheta),
\end{equation*}
and set 
\[\hat\btheta_{\text{de}}=\hat\btheta + \hat\bH(\hat\btheta)^\text{pinv}\bm{\xi}_n(\hat\btheta),
\quad
\hat\btheta_{0,\text{de}}=\hat\btheta_0 + \bH_0(\hat\btheta_0)^\text{pinv}\bm{\xi}_n(\hat\btheta_0).
\]

\begin{assumption}\label{ass:debias}
The de-biasing procedure satisfies
\begin{enumerate}[(a)]
 \item 
  $n^{1/2}(\bH(\btheta^*)^\text{pinv}-\bH_0(\btheta^*)^\text{pinv})\bm{\xi}_n(\btheta^*)\toP \bm{0}$ and $n^{1/2}(\hat\bH(\hat\btheta)^\text{pinv}-\bH_0(\hat\btheta)^\text{pinv})\bm{\xi}_n(\hat\btheta)\toP \bm{0}.$
\item There exist $C_n\toP 0$ so that  
$$\|\bH_0(\btheta_1)^\text{pinv}\bm{\xi}_n(\btheta_1)-\bH_0(\btheta_2)^\text{pinv}\bm{\xi}_n(\btheta_2)\|\leq  C_n\|\btheta_1-\btheta_2\|$$ for all $\btheta_1,\btheta_2\in K$. 
\item $R_n=\sup_{\|\btheta-\btheta_0\|\leq \|\hat\btheta_0-\btheta_0\|}\|\bI - \bH_0(\hat\btheta_0)^\text{pinv}\bH_0(\btheta)\|_2\toP 0$.
\end{enumerate}
\end{assumption}
The following theorem shows that under our assumptions the de-biasing procedure reduces bias without affecting the asymptotic normality of the fiducial samples.
\begin{theorem}\label{thm:debias}
    Under the Assumptions~\ref{ass:consistency}, \ref{ass:normality}, \ref{ass:debias}
    \[
    n^{1/2}(\hat\btheta_{\text{de}}-\hat\btheta_{0,\text{de}})\toD N(\bm{0},\bT_\infty^{-1} \bS(\btheta_0)\bT_\infty^{-1\top} ),\]
    \[
      n^{1/2}(\btheta^*_{\text{de}}-\hat\btheta_{\text{de}})\toD N(\bm{0},\bT_\infty^{-1} \bS(\btheta_0)\bT_\infty^{-1\top} ),
    \]
    and
    \[\hat\btheta_{0,\text{de}}-\btheta_0= o_P(\hat\btheta_0-\btheta_0).\]
\end{theorem}


\begin{remark}
 The work in this section deals with the algorithm without a rejection step. However, in some situations, the rejection step improves the estimator's precision. This is caused by the fact that the event $\mathcal U_{\by,0}$ is often approximately ancillary, in which case conditioning on it could potentially decrease asymptotic variance and improve precision. This phenomenon will be the subject of future study. However, in line with the result of this section and common practice in the ABC literature, we keep the $\epsilon$ relatively large and filter out only extreme in the applications of \AutoGFI.
\end{remark}
\section{Tensor Regression}\label{sec:TR}
\subsection{Background}
Technological advances have led to the generation of vast multi-way array data in areas such as genomics \citep{tao2017generalized} and medical imaging \citep{zhou2013tensor}, which can be naturally represented as tensors. Gene-gene and protein-protein interaction networks can be expressed as tensors of order two (i.e., 2D adjacency matrices), while anatomical MRI can be represented as three-mode tensors. Using tensor regression models, the relationship between this complex data and clinical outcomes can be studied.

Tensor decomposition is a useful method for exploring the low-rank structure of a tensor, as the major component is often governed by a small number of latent factors \citep{kolda2009tensor, shang2014generalized}.  Several tensor regression models have been proposed based on tensor decomposition, such as the CANDECOMP/PARAFAC and Tucker decompositions. While many studies have focused on estimating the tensor coefficient and selecting effective regions, few have addressed quantifying the uncertainty of estimates \citep{guo2011tensor,li2018tucker,ou2020sparse,zhou2013tensor,10643066}. Notable exceptions include the Bayesian approaches introduced in \cite{guhaniyogi2017bayesian} and \cite{papadogeorgou2021soft}.

This section applies \AutoGFI\ to the tensor regression model and compares it with the two Bayesian methods mentioned above.  Results from simulation experiments show that \AutoGFI\ can provide a robust estimate of the tensor coefficient and, at the same time, can offer preferable uncertainty quantification.

\subsection{Problem Definition}
Let $Y_i \in \mathbb{R}$ be a response variable and $\bX_i \in \otimes_{d=1}^D\mathbb{R}^{p_d}$ be a tensor predictor of order $D$ for $i = 1, \dots, n$. We consider the Gaussian linear model for the tensor regression,
\begin{equation}\label{basic_model}
    Y_i = \langle \bX_i,\bB \rangle + U_i,  \quad U_i \stackrel{iid}{\sim}  N(0,\sigma^2),
\end{equation}
where the tensor coefficient $\bB$ is a $D\text{-mode}$ tensor with $\prod_{d=1}^Dp_d$ unknown parameters.  Also, $\langle \bX_i,\bB \rangle = \text{vec}(\bX_i)^\top \text{vec}(\bB)$ and $\text{vec}(\bX_i)$ is the vectorization of $\bX_i$. Our goal is to provide robust estimation and uncertainty quantification for $\bB$.




As the dimensionality of $\bB$ usually exceeds the sample size, regularization techniques are needed to reduce the number of parameters.  We follow \cite{zhou2013tensor} and assume a rank-$R$ CP decomposition of the tensor coefficient, i.e.
$    \bB = \sum_{r=1}^R \bbeta^{(r)}_1\circ \cdots\circ \bbeta_D^{(r)},$
where $\bbeta_d^{(r)} \in \mathbb{R}^{p_d}$ for $r = 1,\dots, R$ and $d = 1, \dots, D$. With this, model \eqref{basic_model} becomes
\begin{equation}
\begin{aligned}
     Y_i &= \langle \bX_i,\bB \rangle + U_i 
       = \langle \bX_i,\sum_{r=1}^R \bbeta^{(r)}_1\circ \cdots\circ \bbeta_D^{(r)} \rangle + U_i \\
    &= \sum_{r=1}^{R}\text{vec}(\bX_i)^\top \bbeta_D^{(r)}\otimes\cdots\otimes \bbeta_1^{(r)}+ U_i,
\end{aligned}
\label{tensor-model} 
\end{equation}
where $U_i \stackrel{iid}{\sim}  N(0,\sigma^2)$. In the above, the coefficient $\bB$ is determined by the factors $\{\bbeta_d^r\}_{d=1\dots D}^{r =1 \dots R}$. Therefore, the number of unknown parameters of $\bB$ decreases from $\prod_{d=1}^Dp_d$ to $R\sum_{d=1}^Dp_d$. 

\subsection{\AutoGFI\ for Tensor Regression}
Model (\ref{tensor-model}) can be re-expressed as
$Y_i = G(\bX_i,\btheta) +  U_i$
with $U_i \stackrel{iid}{\sim}  N(0,\sigma^2)$ and
$G(\bX_i,\btheta) = \langle \bX_i,\sum_{r=1}^R \bbeta^{(r)}_1\circ \cdots\circ \bbeta_D^{(r)} \rangle.$
The semi-metric $\rho$ is then
\begin{equation}\label{tr-rho}
\begin{aligned}
&\rho(\bY,G(\bX,\btheta) + \bU^*) \\
= &\sum_{i=1}^n(Y_i -\langle \bX_i,\sum_{r=1}^R \bbeta^{(r)}_1\circ \cdots\circ \bbeta_D^{(r)}\rangle- U_i^*)^2 ,
\end{aligned}
\end{equation}
where $\bY = (Y_1,\ldots,Y_n)$, $\bX = (\bX_1,\ldots,\bX_n)$ and $\bU^*$ is an independent copy of $\bU$.

Although the low-rank structure assumption reduces the number of parameters significantly, further regularization is required to ensure the number of parameters is less than the number of observations.  
Here we employ an $\ell_1$ penalty to introduce sparsity into the model:
$\lambda(\bbeta) = \lambda \sum_{d,r}\|\bbeta_{d}^{(r)}\|_1$.

When the noise scale $\sigma$ is known, plugging the above $\rho$ and $\lambda$ in the \AutoGFI algorithm, one can generate debiased fiducial samples for the unknown parameters $\btheta = \{\bbeta_d^r\}_{d=1\dots D}^{r=1\dots R}$. Thus, Steps 2 and 3 of Algorithm~\ref{alg:AutoGFI} for tensor regression become:
\begin{enumerate}
    \item[2.] Solve 
    \begin{gather*}
    \btheta^* 
    =  \argmin\limits_{\btheta} \sum_{i=1}^n(Y_i -\langle \bX_i,\sum_{r=1}^R \bbeta^{(r)}_1\circ \cdots\circ \bbeta_D^{(r)}\rangle- U_i^*)^2  + \lambda\sum_{d,r}\|\bbeta_{d}^{(r)}\|_{1}
    \end{gather*}
    with the block relaxation algorithm proposed in \cite{zhou2013tensor}. 
    \item[3.] Debias only the non-zero coordinates as described in Section~\ref{sec:debias}. 
\end{enumerate}
In practice, when $\sigma$ is unknown, we first obtain its MLE $\hat{\sigma}$ before generating the fiducial samples and then replace $\sigma$ by $\hat{\sigma}$ in the algorithm.

A fiducial sample of $m$ copies of $\{\bbeta_d^{r}\}_{d=1\dots D}^{r=1\dots R}$ can be generated by repeating the above algorithm. Each copy forms a fiducial sample of $\bB$, denoted by $\bB^*$. The entries of $\bB$ can then be estimated by taking the element-wise mean or median of the $\bB^*$s. Additionally, the $(1-\alpha)$ confidence interval can be constructed by using the percentiles of the sample.

\subsection{Empirical Performance}
This subsection evaluates the practical performance of \AutoGFI by comparing it to two Bayesian methods in \cite{guhaniyogi2017bayesian} and \cite{papadogeorgou2021soft}. The former utilizes CP decomposition for dimension reduction and assumes a multiway shrinkage prior in the model, while the latter softens the CP decomposition by introducing entry-specific variability to the row contributions. 
We refer to these two methods as Bayesian-hard and Bayesian-soft, respectively.

\begin{figure*}
\centering
\includegraphics[width = 0.8\textwidth]{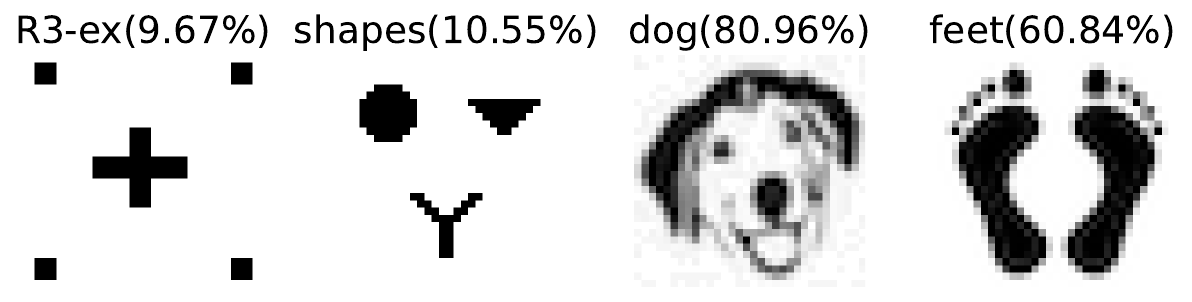}
\caption{\label{fig:TR_B}Four $32 \times 32$ 2D images $\bB$ used in the tensor regression simulation study. Their sparsity levels, defined as \% of non-zero pixels, are displayed in parentheses.}
\end{figure*}

Throughout the simulation study, we set 
$\bX_i$ as a $32 \times 32$ 2D matrix with standard normal entries and the corresponding coefficient $\bB$ as an image varying from low-rank to no low-rank structure, with different degrees of sparsity. Setting $\sigma^2=0.5$, we generated 100 replicated datasets with $n=400$ according to 
(\ref{basic_model}).
The images $\bB$ considered are shown in Figure~\ref{fig:TR_B}. 

The methods are evaluated by their pixel-wise average biases and root-mean-squared errors (rMSEs), as well as frequentist coverages (and widths) of 90\%, 95\%, and 99\% confidence/credible intervals. By default, we set $R = 10$ for \AutoGFI\ and Bayesian-hard and $R = 3$ for Bayesian-soft as recommended by \cite{papadogeorgou2021soft}. 
The regularization parameter $\lambda$ in \AutoGFI\ was chosen by 10-fold cross-validation. 
The accepting threshold $\epsilon$ was automatically chosen as $Q_3 + 1.5(Q_3-Q_1)$ to remove potential outliers of the losses $\rho(\bY, G(\bX,\btheta_\text{de}^*)+\bm{U}^*)$, where the $Q_1$ and $Q_3$ are, respectively, the first and third quartiles of these losses. 
We chose 0.05 as the threshold constant $c$ for debiasing based on our experiment results, which indicated that the singular values of $\hessian$ experienced a significant drop near $0.05\,\zeta_1(\hessian)$.


Table~\ref{tab:tr-rmse} reports the relative performance of the methods on point estimation, which shows that \AutoGFI\ outperformed the two Bayesian methods in both bias and rMSE.


\begin{table}
    \caption{Average biases and rMSEs of the three methods compared in tensor regression. \label{tab:tr-rmse}}
\begin{center}
\resizebox{!}{\height}{%
    \begin{tabular}{llccc} 
    \hline
         &     &  AutoGFI &  Bayesian-hard &  Bayesian-soft\\
    \hline
    \multirow{2}{*}{R3-ex} & bias  &           0.0024 &           0.0141 &           0.0372 \\
     & rMSE  &           0.0079 &           0.0198 &           0.1155 \\
\cline{1-5}
\multirow{2}{*}{shapes} & bias  &           0.0384 &           0.0675 &           0.1431 \\
     & rMSE  &           0.0730 &           0.1008 &           0.3150 \\
\cline{1-5}
\multirow{2}{*}{dog} & bias &    0.1130 &           0.1310 &           0.2260 \\
     & rMSE  &           0.1561 &           0.1736 &           0.3235 \\
\cline{1-5}
\multirow{2}{*}{feet} & bias  &           0.1039 &           0.1226 &           0.2957 \\
     & rMSE  &           0.1469 &           0.1625 &           0.4321 \\
     \hline
    \end{tabular}}
\end{center}
\end{table}

Table~\ref{tab:tr-uq} compares the empirical frequentist coverages and widths of 90\%, 95\%, and 99\% confidence/credible intervals for truly zero and truly non-zero pixel entries in $\bB$. For non-zero entries, when the low-rank assumption is true (R3-ex), \AutoGFI Bayesian-hard performed similarly, with coverages close to the target level.  When $\bB$ is approximately low-rank, \AutoGFI outperformed other methods, with coverage levels much closer to the target levels and much higher than the others when sharing the same order of interval widths. For truly zero elements, \AutoGFI provided higher coverages (and very often smaller interval widths), indicating that they are better at distinguishing significant regions from zeros. This is because shrinkage causes a significant portion of the relevant coordinates in the fiducial sample to be exactly zero.

\begin{table}
\caption{Empirical frequentist coverages and widths (in parentheses) of 90\%, 95\%, and 99\% confidence/credible intervals among truly zero and truly non-zero pixel entries for different methods in tensor regression.  \label{tab:tr-uq} } 
\begin{center}
\resizebox{!}{0.9\height}{%
\begin{tabular}{lllccc}\hline

     &            &     &  AutoGFI &  Bayesian-hard &  Bayesian-soft\\\hline
\multirow{6}{*}{\rotatebox[origin=c]{90}{R3-ex}} & \multirow{3}{*}{\rotatebox[origin=c]{90}{non-zero}} & 90\% &    88.02\% (0.0730) &  90.77\% (0.1032) &   88.03\% (0.3090) \\
     &          & 95\% &  93.87\% (0.0869) &  95.33\% (0.1231) &  91.32\% (0.3725) \\
     &          & 99\% &  98.49\% (0.1131) &   98.91\% (0.1610) &  93.42\% (0.5004) \\
\cline{2-6}
     & \multirow{3}{*}{\rotatebox[origin=c]{90}{zero}} & 90\%  &   99.98\% (0.0030) &  94.99\% (0.0649) &  97.29\% (0.1648) \\
     &            & 95\% &  99.99\% (0.0046) &   97.84\% (0.0790) &  98.35\% (0.2105) \\
     &            & 99\% &  99.99\% (0.0094) &  99.64\% (0.1074) &  99.05\% (0.3075) \\
\cline{1-6}
\multirow{6}{*}{\rotatebox[origin=c]{90}{shapes}} &\multirow{3}{*}{\rotatebox[origin=c]{90}{non-zero}} & 90\% &  89.48\% (0.4225) &  77.16\% (0.4282) &  41.51\% (0.4036) \\
     &          & 95\% &  94.15\% (0.5075) &  84.33\% (0.5072) &  45.44\% (0.4913) \\
     &          & 99\% &  97.83\% (0.6597) &  91.77\% (0.6485) &  49.57\% (0.6721) \\
\cline{2-6}
     & \multirow{3}{*}{\rotatebox[origin=c]{90}{zero}} & 90\% &  98.52\% (0.1422) &  94.92\% (0.2889) &  92.65\% (0.3189) \\
     &            & 95\% &  99.16\% (0.1867) &  97.22\% (0.3464) &  93.58\% (0.3981) \\
     &            & 99\% &  99.79\% (0.2827) &  98.99\% (0.4541) &  94.24\% (0.5683) \\
\cline{1-6}
\multirow{6}{*}{\rotatebox[origin=c]{90}{dog}} & \multirow{3}{*}{\rotatebox[origin=c]{90}{non-zero}} & 90\% &  90.19\% (0.5138) &  82.98\% (0.4757) &  70.32\% (0.5806) \\
     &            & 95\% &  94.29\% (0.6313) &   88.58\% (0.5680) &  75.86\% (0.6993) \\
     &            & 99\% &  98.35\% (0.8853) &  94.51\% (0.7409) &  84.72\% (0.9405) \\
\cline{2-6}
     & \multirow{3}{*}{\rotatebox[origin=c]{90}{zero}} & 90\% &  97.09\% (0.3795) &  90.96\% (0.4394) &  87.71\% (0.5281) \\
     &            & 95\% &  99.39\% (0.4882) &  95.32\% (0.5267) &  90.78\% (0.6405) \\
     &            & 99\% &  100.0\% (0.7434) &  98.81\% (0.6923) &   94.23\% (0.8720) \\
\cline{1-6}
\multirow{6}{*}{\rotatebox[origin=c]{90}{feet}} & \multirow{3}{*}{\rotatebox[origin=c]{90}{non-zero}} & 90\% &  84.35\% (0.4654) &  79.64\% (0.4604) &  50.57\% (0.4993) \\
     &          & 95\% &  90.96\% (0.5709) &   86.00\% (0.5484) &  56.01\% (0.5991) \\
     &          & 99\% &  97.62\% (0.7945) &  93.45\% (0.7102) &  64.39\% (0.7989) \\
\cline{2-6}
     & \multirow{3}{*}{\rotatebox[origin=c]{90}{zero}} & 90\% &  96.13\% (0.3088) &   91.97\% (0.4060) &  79.68\% (0.4175) \\
     &            & 95\% &  98.62\% (0.4004) &  95.79\% (0.4864) &  82.39\% (0.5051) \\
     &            & 99\% &  99.95\% (0.6155) &  98.64\% (0.6375) &  84.53\% (0.6842)\\
     \hline
\end{tabular}}%
\end{center}
\end{table}

\section{Matrix Completion}\label{sec:MC}
\subsection{Background}
Matrix completion is a fundamental problem in machine learning, encountered in many applications, with the Netflix movie rating challenge \citep{bennett2007netflix} being perhaps the most well-known example. Here the dataset is a large movie rating matrix consisting of 17770 movies (columns) and 480189 customers (rows), with less than 1\% of the data matrix (customer-movie pairs) observed. The challenge participants were asked to develop methods to impute the unobserved movie ratings, which is an ill-specified problem and requires additional constraints on the unknown full matrix to make it well-defined. Rank constraints are the most popular choices, with many solutions assuming the full matrix is of low rank. This low-rank assumption is well-empirically motivated in many applications; for example, in the Netflix challenge, it corresponds to the belief that users' movie ratings are based on a few factors.  Other applications areas for matrix completion include computer vision \citep{chen2004recovering}, medical imaging \citep{haldar2010spatiotemporal}, collaborative filtering \citep{rennie2005fast}, and others.  This section applies \AutoGFI\ to this matrix completion problem.

\subsection{Problem Definition}
Let $\bM$ be a real matrix of size $m\times n$. Only a small fraction of the elements $M_{ij}$'s of $\bM$ are observed. Denote the index set of the observed elements as $\Omega=\{1,\ldots,m\}\times \{1,\ldots,n\}$; that is, $(i,j)\in \Omega$ if $M_{ij}$ is observed. For simplicity, we collect all the observed elements in an observed matrix $\bY$, through a projection function $f(\bM)$ of $\bM$; that is, 
\[
Y_{ij}=
\begin{cases}
M_{ij} \quad \mbox{if } (i,j)\in \Omega \\
0 \quad \mbox{otherwise}. 
\end{cases}
\]
The goal is to, given $\bY$, estimate the unknown elements of $\bM$.
Under the low-rank assumption and in the absence of noise, the matrix completion problem can be formulated as a rank-minimization problem: 
\begin{equation}
\label{eqn:rank}    
\min_{\bM} \text{rank}(\bM) \quad \textrm{s.t. } \bY = \bm{f}(\bM).
\end{equation}
Although~(\ref{eqn:rank}) guarantees the exact recovery of $\bM$ under some regularity conditions \citep{candes2011robust, candes2009exact}, it is a non-convex and NP-hard problem \citep{srebro2003weighted} so no known polynomial-time solutions exist.  Various computationally feasible reformulations have been proposed to overcome this limitation, e.g., \cite{candes2010matrix, chen2015fast, gross2011recovering, keshavan2010matrix, koltchinskii2011nuclear, negahban2012restricted}.

This section focuses on a reformulation that allows for the observed entries to be corrupted by additive noise and assumes the true matrix $\bM$ can be factorized into two rank-$R$ factor matrices, i.e. $\bM=\bA \bB^\top$. In practice, the matrix rank $R$ can often be reliably estimated in a data-dependent manner \citep[e.g.,][]{chen2019inference}.
The solution to this reformulation is:
\begin{equation}
\label{eqn:mc-nonconvex}
 \min_{\bA,\bB} \|\bY-\bm{f}(\bm{AB^\top})\|_{F}^{2} + \lambda\|\bA\|^2_{F} +\lambda\|\bB\|_{F}^2,
\end{equation}
where $\|\cdot\|_{F}$ is the Frobenius norm and $\lambda$ is the regularization parameter.  This reformulation has been studied, for example, in \cite{chen2019inference, chen2020noisy, yuchi2022bayesian}.

\subsection{\AutoGFI\ for Matrix Completion}\label{sec:MCmethodology}


With the assumptions behind~(\ref{eqn:mc-nonconvex}), the noisy matrix observation model is
\begin{equation}\label{MC-model}
    Y_{ij} = (\bA\bB^\top)_{ij} + U_{ij} \quad \text{ for all } (i,j)\in \Omega,
\end{equation}
where $U_{ij}$ denotes independently generated normal noise at the location $(i, j)$. We assume the data are missing uniformly at random.  That is, each index $(i, j)$ is included in $\Omega$ independently with the same probability.  Therefore, the data generating equation of \eqref{MC-model} is
$    \bY= \bm{f}(\bA\bB^\top) +  \bU,$
where $\bU$ is a matrix with $U_{ij} \stackrel{iid}{\sim}  N(0,\sigma^2) \text{ for all } (i,j)\in \Omega$ and $U_{ij} = 0 \text{ for all } (i,j)\notin \Omega$. The semi-metric $\rho$ is 
\begin{equation}
\label{mc-rho}
    \rho(\bY,\bm{G}(\btheta)+\bU^*)=\|\bY-\bm{f}(\bA\bB^{T})-\bU^*\|_{F}^{2},
\end{equation}
where $\btheta = (\bA, \bB)$. 
and with penalty chosen as $\lambda(\btheta) = \lambda\|\bA\|_{F}^{2}+\lambda\|\bB\|_{F}^{2}$. 
This leads to an optimization problem that is similar to (\ref{eqn:mc-nonconvex}) and can be efficiently solved by the 2-stage algorithm of \cite{chen2019inference}. With these, the fiducial samples of $\bA$ and $\bB$ can be generated by Algorithm~\ref{alg:AutoGFI}, where Steps~1 to~3 become:
\begin{enumerate}
    \item Generate $U_{ij}^* \sim  N(0,\sigma^2)$ for $(i,j) \in \Omega$ and leave the other entries as 0.
    \item Solve 
    \begin{equation*}
    \btheta^* = (\bA^*, \bB^*)
              = \argmin_{\bA, \bB} \|\bY-f(\bA\bB^{T})- \bU^{*}\|_{F}^{2}+\lambda(\|\bA\|_{F}^{2}+\|\bB\|_{F}^{2})
    \end{equation*}
    using the 2-stage algorithm of \cite{chen2019inference}.
\item Debias the non-zero coordinates as described in Section~\ref{sec:debias} using $\rho$ defined in (\ref{mc-rho}) and $\bm{\xi}$ the gradient of the penalty term.         
\end{enumerate}
When $\sigma$ is unknown, we first estimate it using the median absolute deviation (MAD) estimator before generating the fiducial sample and then replace $\sigma$ with the resulting MAD estimate.

Using the procedure introduced above, one can obtain fiducial samples of $\bA$ and $\bB$ and hence $\bM=\bA\bB^\top$.  From these samples, the point estimates and confidence intervals of the missing entries of $\bM$ can be obtained.



\subsection{Empirical Performance}
This section compares the empirical performance of \AutoGFI\ for matrix completion with the frequentist method proposed in \cite{chen2019inference}, referred to as Freq-MC below, and the
Bayesian method BayeSMG proposed in \cite{yuchi2022bayesian}.


Similar to the experimental setting of \cite{chen2019inference}, we generated a rank-$R$ matrix $\bM = \bA\bB^\top$, where $\bA,\bB \in \mathrm{R}^{n \times R}$ are random orthonormal matrices. We added noise with $\sigma = 0.001$ to obtain $\bY = \bM+\bU$, where $U_{ij} \sim N(0,\sigma^2)$ are i.i.d. Each entry of $\bY$ is observed with probability $p$ independently. We considered two values for $n$: (500, 1000), two values for $R$: (2, 5), and two values for $p$: (20\%, 40\%), resulting in eight simulation settings. 

The methods are evaluated by estimation errors measured in Frobenuis norm and frequentist coverages (and widths) of 90\%, 95\%, and 99\% confidence/credible intervals.  We used 10-fold cross-validation to choose the penalty term $\lambda$ for \AutoGFI\ and set $\lambda=2.5\sigma\sqrt{np}$ for Freq-MC as suggested by the authors. Same as tensor regression, the accepting threshold $\epsilon$ was chosen as $Q_3 + 1.5(Q_3-Q_1)$, where the $Q_1$ and $Q_3$ are the, respectively, first and third quartiles of the losses $\rho(\bY, G(\btheta_{\text{de}}^*)+\bU^*)$.  
The threshold constant $c$ used in debiasing was chosen as 0.05, the same as tensor regression.

For each simulation setting, we randomly generated 100 matrices with missing entries, and for each matrix, we applied the above three methods to estimate its missing entries and construct confidence intervals.  Table~\ref{tab:MC-pointEst} reports the 
estimation errors of the methods, while Tables~\ref{tab:MC-uq} shows the empirical coverage rates.
Notice that \AutoGFI\ gave the lowest estimation errors and produced comparable confidence interval coverages when compared to other tailor-made methods.

\begin{table}
\caption{\label{tab:MC-pointEst} Average estimation errors of the three methods compared in matrix completion.}
\begin{center}
\resizebox{!}{\height}{%
    \begin{tabular}{llccc}\hline
           &       &AutoGFI & BayeSMG &  Freq-MC\\\hline
\multirow{2}{*}{$n=500, \: R=2$} & $p=0.2$ &           0.1035 &   0.1449 &   0.1125\\
           & $p=0.4$ &           0.0724 &   0.1002 &   0.0738\\
\cline{1-5}
\multirow{2}{*}{$n=500, \: R=5$} & $p=0.2$&           0.1707 &   0.2335 &   0.1944\\
           & $p=0.4$ &           0.1156 &   0.1611 &   0.1207\\
\cline{1-5}
\multirow{2}{*}{$n=1000, \: R=2$} & $p=0.2$ &           0.1469 &   0.2027 &   0.1500\\
           & $p=0.4$ &           0.1040 &   0.1417 &   0.1020\\
\cline{1-5}
\multirow{2}{*}{$n=1000, \: R=5$} & $p=0.2$ &           0.2373 &   0.3214 &   0.2457\\
           & $p=0.4$ &           0.1629 &   0.2248 &   0.1625 \\
\hline
\end{tabular}}
\end{center}
\end{table}

\begin{table}
\caption{\label{tab:MC-uq} Empirical frequentist coverages and widths (in parentheses) of 90\%, 95\%, and 99\% confidence/credible intervals for the missing entries in matrix completion.}
\begin{center}
\resizebox{!}{0.9\height}{%
    \begin{tabular}{lllccc}\hline
           &  &     & AutoGFI &           BayeSMG &           Freq-MC \\\hline
\multirow{6}{*}{\rotatebox[origin=c]{90}{$n=500, \: R=2$}} & \multirow{3}{*}{\rotatebox[origin=c]{90}{$p=0.2$}} & 90\% &  89.08\% (0.0006) &  89.82\% (0.0006) &  86.27\% (0.0006) \\
           &       & 95\% &  94.30\% (0.0007) &  94.85\% (0.0008) &  92.18\% (0.0007) \\
           &       & 99\% &   98.59\% (0.001) &   98.90\% (0.001) &   97.73\% (0.001) \\
\cline{2-6}
           & \multirow{3}{*}{\rotatebox[origin=c]{90}{$p=0.4$}} & 90\% &  89.31\% (0.0004) &  89.92\% (0.0004) &  88.44\% (0.0004) \\
           &       & 95\% &  94.46\% (0.0005) &   94.93\% (0.0005) &  93.89\% (0.0005) \\
           &       & 99\% &  98.67\% (0.0007) &  98.96\% (0.0007) &   98.57\% (0.0007) \\
\cline{1-6}
\multirow{6}{*}{\rotatebox[origin=c]{90}{$n=500, \: R=5$}} & \multirow{3}{*}{\rotatebox[origin=c]{90}{$p=0.2$}} & 90\% &   87.73\% (0.001) &  89.78\% (0.0011) &   81.94\% (0.001) \\
           &       & 95\% &  93.34\% (0.0012) &   94.83\% (0.0013) &   88.78\% (0.0012) \\
           &       & 99\% &  98.24\% (0.0016) &  98.90\% (0.0017) &    96.04\% (0.0015) \\
\cline{2-6}
           & \multirow{3}{*}{\rotatebox[origin=c]{90}{$p=0.4$}} & 90\% &   88.64\% (0.0007) &  89.90\% (0.0007) &  86.69\% (0.0007) \\
           &       & 95\% &  94.00\% (0.0008) &  94.92\% (0.0009) &  92.62\% (0.0008) \\
           &       & 99\% &  98.50\% (0.0011) &  98.96\% (0.0011) &  98.07\% (0.0011) \\
\cline{1-6}
\multirow{6}{*}{\rotatebox[origin=c]{90}{$n=1000, \: R=2$}} & \multirow{3}{*}{\rotatebox[origin=c]{90}{$p=0.2$}} & 90\% &  89.54\% (0.0004) &  89.79\% (0.0004) &  88.15\% (0.0004) \\
           &       & 95\% &  94.57\% (0.0005) &  94.81\% (0.0005) &  93.62\% (0.0005) \\
           &       & 99\% &  98.70\% (0.0007) &  98.89\% (0.0007) &  98.43\% (0.0007) \\
\cline{2-6}
           & \multirow{3}{*}{\rotatebox[origin=c]{90}{$p=0.4$}} & 90\% &  89.57\% (0.0003) &  90.10\% (0.0003) &  89.46\% (0.0003) \\
           &       & 95\% &  94.67\% (0.0004) &  95.05\% (0.0004) &  94.62\% (0.0004) \\
           &       & 99\% &  98.76\% (0.0005) &  98.95\% (0.0005) &  98.83\% (0.0005) \\
\cline{1-6}
\multirow{6}{*}{\rotatebox[origin=c]{90}{$n=1000, \: R=5$}} & \multirow{3}{*}{\rotatebox[origin=c]{90}{$p=0.2$}} & 90\% &   89.01\% (0.0007) &  89.84\% (0.0007) &  86.44\% (0.0007) \\
           &       & 95\% &  94.22\% (0.0009) &  94.86\% (0.0009) &  92.38\% (0.0008) \\
           &       & 99\% &  98.60\% (0.0011) &  98.90\% (0.0011) &  97.94\% (0.0011) \\
\cline{2-6}
           & \multirow{3}{*}{\rotatebox[origin=c]{90}{$p=0.4$}} & 90\% &  89.66\% (0.0005) &  90.06\% (0.0005) &  88.96\% (0.0005) \\
           &       & 95\% &  94.68\% (0.0006) &  95.01\% (0.0006) &  94.26\% (0.0006) \\
           &       & 99\% &   98.71\% (0.0008) &  98.97\% (0.0008) &  98.69\% (0.0008) \\
           \hline
\end{tabular}}%
\end{center}
\end{table}
\section{Regression with Network Cohesion}\label{sec:NR}
\subsection{Background}

As modern communication technology advances, network data are becoming increasingly popular. One longstanding problem is community detection, such as identifying friendship circles in a social network. Also, combining information from node features and the network structure has gained interest among researchers, such as using node covariates to assist in inferring the network structure  
\citep[e.g.,][]{binkiewicz2017covariate, su2020network, zhang2016community}.

One can also leverage network structures to assist inference on the node covariates, which is the main focus of this section.  We consider a linear regression model with observations connected in a network, where each node is associated with some node covariates and a response variable of interest. To incorporate network effects into traditional predictive models,  \cite{li2019prediction} proposed an RNC (regression with network cohesion) estimator that uses a graph-based regularization to assert similar individual effects for those who are connected in the network. They showed that by adding the network cohesion, the out-of-sample prediction error was significantly improved. This idea of adding graph-regularization has also been applied to many other problems, such as graph-regularized matrix completion \citep{ma2011recommender,NIPS2015_5938}.

This section demonstrates the use of \AutoGFI\ in the RNC problem to provide uncertainty quantification of the model parameters. We note that when the noise is additive, \AutoGFI\ can be straightforwardly extended to other graph-regularized methods.

\subsection{Problem Definition}
In linear regression with network cohesion \citep{li2019prediction}, the model is
\begin{equation}
	\label{eqn:NRmodel}
	\bY = \bX\bbeta + \balpha + \bU,
\end{equation}
where $\bY=(Y_1, \ldots, Y_n)^T$ is the response vector, $\bX$ is an $n\times p$ design matrix, $\balpha$ is an $n\times 1$ vector of individual effects, $\bbeta$ is a $p \times 1$ vector of fixed effects covariates, and $\bU\sim N(\bm{0},\sigma^2 \bm{I}_n)$ is the error term. 

We assume these $n$ samples are connected in a network $\mathcal{G}=(V, E)$, where $V=\{1,\ldots,n\}$ is the node set, and $E\subset V\times V$ is the edge set. The Laplacian of $\mathcal{G}$ is defined as $\bm{L}=\bm{D}-\bm{A}$ where $\bm{A}$ is the adjacency matrix and $\bm{D}=\text{diag}(d_1,\ldots,d_n)$ with $d_i$ being the degree of the $i$-th node.  The RNC estimator is defined as the minimizer of
\begin{equation}
	\label{eqn:NRobj}
	L(\balpha, \bbeta) = \Vert \bY - \bX\bbeta - \balpha\Vert^2 +\lambda \balpha^T\bL\balpha,
\end{equation}
where $\lambda>0$ is a tuning parameter selected by cross-validation.

In general, the $n+p$ parameters ($\balpha$ and $\bbeta$) cannot be estimated from $n$ observations without further assumptions. In this network regression setting, we assume that the design matrix $\bX$ is centered and has full column rank. \cite{li2019prediction} proved that when the network contains additional information beyond what is contained in $\bX$, the RNC estimator always exists. We will make the same assumption here.

\cite{li2019prediction} also derived the asymptotic bounds for the bias and the variance of the RNC estimator. In particular, the bias term depends on $\bm{L}\balpha$, and the norm of bias grows with $\Vert \bm{L}\balpha\Vert$. Under the condition that $\Vert \bm{L}\balpha\Vert=0$, the RNC estimator is unbiased. This occurs when the individual effect on node $i$ is simply the average of those of its neighboring nodes; otherwise, there is bias involved.


\subsection{\AutoGFI\ for Network Regression}\label{RNCMethod}

First, the data generating equation of network regression is given in~\eqref{eqn:NRmodel}. Using our notations from Sections~\ref{sec:GFIBackground} and~\ref{sec:NewGFI}, we can express it in the form of $\bY =\bG(\bX,\btheta)+\bU$ with $\bG(\bX,\bm{\btheta}) = \bX\bbeta + \balpha$. The unknown parameters are $\btheta=(\balpha,\bbeta)$, and the random component is $\bU$.
Also, the semi-metric $\rho$ for this model is
\begin{equation}
\label{eqn:NRGFIdef}
 \rho(\bY, \bm{G}(\bX,\btheta)+ \bU^*) = \Vert \bY - \bX\bbeta - \balpha - \bU^*\Vert^2,
\end{equation}
where $\bU^*$ is an independent copy of $\bU$. According to \eqref{eqn:NRobj}, the penalty term is 
\begin{equation}
\label{eqn:NRPenalty}
\lambda(\btheta) = \lambda \balpha^T\bL\balpha.
\end{equation}
Different from the penalty terms used in Sections~\ref{sec:TR} and~\ref{sec:MC}, \eqref{eqn:NRPenalty} only penalizes the part of $\balpha$ that lies in the column space of $\bL$. The part of $\balpha$ in the null space has no contribution to the penalty term. Therefore, we only need to debias the part in the column space of $\bL$, i.e., we only debias what was penalized.

The optimization problem in Step 2 of \AutoGFI\ can be solved by the algorithm in \cite{li2019prediction}.  To debias the penalized parameters, we separate $\balpha$ as
$\balpha = (\bI-\bm{P_L})\balpha + \bL^{-1/2}\bmeta$
and transform \eqref{eqn:NRGFIdef} to
\begin{equation}
\rho_{\bmeta}(\bY, \bG(\bX,\btheta)+\bU^*) =  \Vert \bY - \bX\bbeta - (\bI-\bm{P_L})\balpha - \bL^{-1/2}\bmeta - \bU^* \Vert^2,
\label{eqn:NRmodel_eta}    
\end{equation}
where $\bm{P_L}$ is the projection matrix of $\bL$ and $\bmeta = \bL^{1/2}\balpha$. 

From~\eqref{eqn:NRmodel_eta}, we can see that $\bmeta$ is the term that we penalize. Thus, we first only debias $\bmeta$, then use $d(\bmeta)$ to rebuild a debiased $\balpha$, and finally refit the model to debias $\bbeta$.  
Therefore, Steps~2 and~3 of Algorithm~\ref{alg:AutoGFI} for network regression become:
\begin{enumerate}
    \item[2.] Solve 
    $$\btheta^*=(\balpha^*, \bbeta^*) = \argmin\limits_{\balpha,\bbeta} \Vert \bY - \bX\bbeta - \balpha - \bU^* \Vert^2 +\lambda \balpha^TL\balpha$$ 
    with the algorithm in \cite{li2019prediction}.
    \item[3.] Let $\bmeta^* = \bL^{{1}/{2}}\balpha^*$ and plug it in \eqref{eqn:NRmodel_eta} so that
    $$
        \rho_{\bmeta}(\bY, \bG(\bX,\btheta)+\bU^*) = \Vert \bY  - \bX\bbeta^* - (\bI-\bm{P_L})\balpha^* -\bL^{-1/2}\bmeta^* - \bU^*\Vert^2.
    $$
    \begin{enumerate}
    \item Debias the non-zero coordinates of $\bmeta^*$ by $\bmeta^*_{\text{de}} = \bmeta^*+\bm{H}(\bmeta^*)^{\text{pinv}}\bm{\xi}(\bmeta^*)$. 
    \item Calculate the debiased $\balpha^*$ as $\bm{\alpha^*}_{\text{de}} = \bm{P_L}\balpha^* + \bL^{-1/2}\bmeta^*_{\text{de}}$.
    \item Obtain the debiased $\bbeta^*$ by refitting the model, i.e.,
    $$\bbeta^*_{\text{de}} = (\bX^\top \bX)^{-1}\bX^\top(\bY-\bL^{-1/2}\bmeta^*_\text{de}).$$ 
    \end{enumerate}
    Set $\btheta_{\text{de}}^* = (\bbeta_{\text{de}}^*,\balpha_{\text{de}}^*)$.
\end{enumerate}
In the above, as the Laplacian matrix $\bL$ is not of full rank, the generalized inverse of $\bL^{1/2}$ is used instead. This enables the construction of a $(1-\alpha)$ fiducial confidence interval for $\bbeta$ using the $\alpha/2$ and $1-\alpha/2$ quantiles of $\bbeta^*_{\text{de}}$. 
Estimating the error scale $\sigma$ is also necessary for this problem. In this study, we use cross-validation and the estimated mean squared prediction error (MSPE) to estimate it. Specifically, before generating the fiducial sample, we apply 10-fold cross-validation to split the data into training and validation sets. For each fold, we estimate the model on the training set and make predictions on the validation set. We compute the MSPE as follows:
$\text{MSPE}_k = \frac{1}{n_{\text{val},k}} \|\bY_{\text{val},k} - \hat{\bY}_{\text{val},k}\|^2$ where the subscript denotes the $k$-th validation set.
The overall estimate $\hat{\sigma}^2$ is then obtained by averaging the $\text{MSPE}_k$ values across all folds. Finally, $\sigma$ is replaced by $\hat{\sigma}$ in the algorithm.

\subsection{Empirical Performance}

We evaluated the practical performance of \AutoGFI\ using a stochastic block model with 3 blocks and $n = 300$ nodes. The probability of a connection between two nodes in the same block was $p_w$, and the probability of a connection between nodes in different blocks was $p_b$. The adjacency matrix $\bm{A}\in\{0,1\}^{n\times n}$ was generated independently with $A_{ij}=A_{ji}$ and $A_{ii}=0$ from the probability matrix $\bm{P}=\bm{B}\otimes \bm{J}_{n/3}$, where $\bm{B}$ is given by 
\begin{equation*}
\bm{B} =
\begin{pmatrix}
p_w & p_b & p_b\\
p_b & p_w & p_b\\
p_b & p_b & p_w
\end{pmatrix}
\end{equation*}
and $\bm{J}_k$ is a $k\times k$ matrix with all ones. 

We generated $\bbeta$ from $N(\bm{1},\bI_p)$, where $p=10$ in our experiments, and the covariate matrix $\bX$ was generated independently from standard normal. Therefore, the columns are uncorrelated and have 0 means. The $\balpha$s were generated independently from a normal distribution with the mean determined by the node's block assignment $N(\eta_k, s^2)$, where $\eta_1=-1, \eta_2=0, \eta_3=1$. Finally, $\bY=\balpha + \bX^T\bbeta + \bU$ with $\bU \sim N(\bm{0},\sigma^2\bI_n)$.



As mentioned before, the bias of the RNC estimator depends on $\Vert \bm{L}\balpha \Vert$. We tested \AutoGFI\ under both unbiased (i.e., $\|\bL\balpha\|=0$) and biased (i.e., $\|\bL\balpha\| > 0$) conditions.  Specifically, we set $p_w=0.2$ and $\sigma=0.5$ and used three choices of $(p_b, s)$: (i) (0, 0) which gives $\|\bL\balpha\|=0$, (ii) (0.01, 0.1) which gives $\|\bL\balpha\|\approx 65$, and (iii) (0.02, 0.1) which gives $\|\bL\balpha\| \approx 108$.




Once the network $\bA$, design matrix $\bX$ and parameters ($\balpha$, $\bbeta$) were generated, we fixed them. We then simulated 500 datasets by re-generating $\bU$ and generating 1000 fiducial samples of ($\balpha^*$, $\bbeta^*$) for each dataset using \AutoGFI. These samples were then used to form confidence intervals.

We compared \AutoGFI\ with two other methods. The first was ordinary linear regression (OLR) without using network information. We used the true $\sigma^2$ and the classical way of constructing confidence intervals. Since the network is correlated with individual effects, ignoring them would cause information to be lost and worsen estimation results. The second method was based on a fixed effects linear model, for which we assumed the true group assignments of nodes are known.  That is, there were 3 known groups, and the individuals within each group shared a common intercept.  When $s\neq 0$, the randomness in $\balpha$ was simply combined into $\bU$, i.e., ${U}_i\sim N(0,s^2+\sigma^2)$.  Note that this second method could be seen as an oracle method since it has access to the usually unknown group assignment. We compared the methods' rMSEs and the coverages and widths of their confidence intervals.  The results are summarized in Table~\ref{tbl:NR1} for different settings.

We can see \AutoGFI outperforms the OLR method and is very similar to the oracle in terms of rMSEs when estimating $\bbeta$s. The coverage of \AutoGFI confidence intervals is also very close to the target level in all cases. As the value of $\Vert\bL\balpha\Vert$ increases, all confidence intervals tend to be wider due to more noise in the data.  Compared to the oracle, the \AutoGFI confidence intervals are often slightly wider because $\sigma^2$ is slightly overestimated, resulting in more dispersed fiducial samples and thus wider confidence intervals. However, under the unbiased case (i.e., $\|\bL\balpha\|=0$), the \AutoGFI confidence intervals are almost identical to the oracle ones. On the other hand, the confidence intervals of OLR have dramatically lower coverage than the target. This shows the necessity of taking the network information into consideration.

\begin{table*}
    \caption{\label{tbl:NR1}Average rMSEs of the parameter estimates and empirical frequentist coverages and widths (in parentheses) of various confidence intervals in network regression.}
    \begin{center}
\resizebox{!}{\height}{%
    \begin{tabular}{lcccc}\hline
    &   &  AutoGFI  & oracle      & OLR \\\hline
    \multirowcell{4}{$p_b=0$, \\ $s=0$, \\$\|\bL\balpha\| = 0$}
     & rMSE & 0.0292 &  0.0290 &  0.0537 \\
     & 90\% &       89.32\% (0.0992) &  89.98\% (0.0981) &    62.9\% (0.098) \\
     & 95\% &       94.8\% (0.118) &   95.38\% (0.117) &  71.46\% (0.1169) \\
     & 99\% &       98.68\% (0.1535) &   99.1\% (0.1542) &   83.68\% (0.154) \\
    \hline
    \multirowcell{4}{$p_b=0.01$, \\ $s=0.1$, \\$\|\bL\balpha\| \approx 65$}
    & rMSE & 0.0298 &  0.0295 &  0.0571 \\
    & 90\% &       90.04\% (0.1055) &   90.0\% (0.1002) &  54.46\% (0.0979) \\
    & 95\% &       94.68\% (0.1255) &  95.04\% (0.1195) &  64.66\% (0.1167) \\
    & 99\% &       98.66\% (0.1632) &  99.12\% (0.1574) &  80.72\% (0.1538) \\
    \hline
    \multirowcell{4}{$p_b=0.02$, \\ $s=0.1$, \\$\|\bL\balpha\| \approx 108$} 
    & rMSE & 0.0318 &  0.0299 &  0.0606 \\
    & 90\% &       91.62\% (0.1157) &  89.58\% (0.1008) &  55.16\% (0.0984) \\
    & 95\% &       95.78\% (0.1377) &  95.04\% (0.1203) &   64.0\% (0.1173) \\
    & 99\% &       99.12\% (0.1791) &  99.12\% (0.1585) &  77.46\% (0.1546) \\\hline
    \end{tabular}}
    \end{center}
\end{table*}

\section{Conclusion}\label{sec:conclude}
In this paper, we discussed the development of GFI and proposed a new form that can be applied to a broad range of high-dimensional and/or nonlinear additive noise problems.  We developed a practical and straightforward-to-implement algorithm, \AutoGFI, for generating fiducial samples of the parameters of interest.  This approach is particularly useful in situations where uncertainty quantification has been difficult or impossible using traditional inference methods.  Numerical results of applying \AutoGFI to three challenging problems demonstrate its highly competitive performance compared to tailor-made competitors. Overall, this paper has shown that GFI is a promising alternative to traditional methods for addressing important and practical inference problems.
To further widen the practical applicability of GFI, a natural and important direction is to extend the \AutoGFI framework to non-additive noise problems, such as generalized linear models or multiplicative noise modeling.  Another direction is to accommodate non-\iid noise scenarios, including the presence of outliers and serial correlation.




\bigskip
\begin{center}
{\large\bf SUPPLEMENTARY MATERIAL}
\end{center}

\section*{Technical Details}\label{sec:app}
This appendix outlines the proofs for Theorems~\ref{thm:consistency} to \ref{thm:debias}.

\begin{proof}[Proof for Theorem~\ref{thm:consistency}.]
 Select $\tau > 0$ 
 and set $C=\min(h(\tau)/2,\tau)$.
 
 By Assumption~\ref{ass:consistency}(c),
\[
 P\left(n^{-1}\sup_{\btheta\in K}\|\sum_{i=1}^n - \nabla^\top_\btheta G(\bX_i,\btheta)  U_i +\bm{\xi}_n(\btheta)\|\geq C\right)\to 0.
\]
Since \eqref{eq:PointEst} is equivalent to
\begin{align*}
 &n^{-1} \sum_{i=1}^n \nabla^\top_\btheta G(\bX_i,\hat\btheta)(G(\bX_i,\btheta_0)-G(\bX_i,\hat\btheta))\\=& 
 n^{-1}\bm{\xi}_n(\hat\btheta)- n^{-1}\sum_{i=1}^n \nabla^\top_\btheta G(\bX_i,\hat\btheta) U_i,
\end{align*}
there is a solution $\hat\btheta\in K$ with probability going to 1 by Assumption~\ref{ass:consistency}(a).
At the same time, if $\hat\btheta\in K$ and  $\|\hat\btheta-\btheta_0\|\geq\tau$, 
then 
\[
 n^{-1}\| \sum_{i=1}^n \nabla^\top_\btheta G(\bX_i,\hat\btheta)(G(\bX_i,\btheta_0)-G(\bX_i,\hat\btheta))\|\geq h(\tau) 
\]
with probability going to 1.
By the union bound
$P(\|\hat\btheta-\btheta_0\|\geq \tau)\to 0$.

The same argument also shows that
$P(\|\btheta^*-\btheta_0\|\geq \tau)\to 0$.

\end{proof}

\begin{proof}[Proof for Theorem~\ref{thm:normality}.]
    The same argument as in the proof of Theorem~\ref{thm:consistency} shows that 
    $\hat\btheta_0\toP \btheta_0$. 
    
    Rewrite \eqref{eq:PointEst} using \eqref{eq:Taylor} to obtain
    \begin{align*}
     &\bT_n(\hat\btheta_0)n^{1/2}(\hat\btheta-\hat\btheta_0) + n^{1/2}\bR_n(\hat\btheta,\hat\btheta_0)\\
     =&
     n^{-1/2} \sum_{i=1}^n \nabla^\top_\btheta G(\bX_i,\hat\btheta)  U_i.
    \end{align*}
    By Theorem~\ref{thm:consistency} $\hat\btheta-\hat\btheta_0\toP 0$ and therefore Assumption~\ref{ass:normality}(b) implies that  $n^{1/2}\bR_n(\hat\btheta,\hat\btheta_0)=o_p(\|n^{1/2}(\hat\btheta-\hat\btheta_0)\|)$. Finally, Assumptions~\ref{ass:normality}(a) and~\ref{ass:consistency}(c) and Slutsky's lemma imply that $n^{1/2}(\hat\btheta-\hat\btheta_0)\toD N(\bm{0},\bT_\infty^{-1} \bS(\btheta_0)\bT_\infty^{-1\top} )$. The second part of the theorem is proved similarly. 
\end{proof}

\begin{proof}[Proof for Theorem~\ref{thm:debias}.]
    Notice that
    \begin{align*}
        n^{1/2}(\btheta^*_{\text{de}}-\hat\btheta_{\text{de}})
        &=n^{1/2}(\btheta^*-\hat\btheta)\\ 
        &+ n^{1/2}\left(\bH(\btheta^*)^{\text{pinv}}-\bH_0(\btheta^*)^{\text{pinv}}\right)\bm{\xi}_n(\btheta^*)\\
        &+n^{1/2}\left(\bH_0(\btheta^*)^\text{pinv}\bm{\xi}_n(\btheta^*)-
       \bH_0(\hat\btheta)^\text{pinv}\bm{\xi}_n(\hat\btheta) \right)\\
        &+n^{1/2}\left(\bH_0(\hat\btheta)^\text{pinv}-\hat\bH(\hat\btheta)^\text{pinv} \right)\bm{\xi}_n(\hat\btheta).
    \end{align*}
    
    The claimed asymptotic normality now follows using Slutsky's lemma, Theorem~\ref{thm:normality} and Assumption~\ref{ass:debias}. The argument for $n^{1/2}(\hat\btheta_{\text{de}}-\hat\btheta_{0,\text{de}})$ is analogous.

    Next, Taylor's theorem and \eqref{eq:BiasEst} imply
    \[
     -\bm{\xi}_n(\hat\btheta_0)=\left(\int_0^1\bH_0(v\hat\btheta_0+(1-v)\btheta_0)\,dv\right)(\hat\btheta_0-\btheta_0).
    \]
    Consequently,
    \[
      \|\hat\btheta_{0,\text{de}}-\btheta_0 \|\leq R_n \,\|\hat\btheta_{0}-\btheta_0\|.
    \]
    This concludes the proof.   
\end{proof}

\bibliographystyle{jasa3}

\bibliography{10.reference}
\end{document}